\documentstyle[graphics,epsf]{lamuphys}
\makeatletter
\let\chapter\hid@chapter
\makeatother

\def\eg{{\em e.g.,}}
\def\ie{{\em i.e.,}}

\def\spose#1{\hbox to 0pt{#1\hss}}
\def\simlt{\mathrel{\spose{\lower 3pt\hbox{$\mathchar"218$}}
     \raise 2.0pt\hbox{$\mathchar"13C$}}}
\def\simgt{\mathrel{\spose{\lower 3pt\hbox{$\mathchar"218$}}
     \raise 2.0pt\hbox{$\mathchar"13E$}}}

\def\deg{\ifmmode {^{\circ}}\else {$^\circ$}\fi}
\def\lya{Ly$\alpha$}
\def\ergsHz{\ifmmode {\rm\,ergs\,s^{-1}\,Hz^{-1}}\else
    ${\rm\,ergs\,s^{-1}\,Hz^{-1}}$\fi}

\def\apj{{ApJ}}
\def\apjl{{ApJ}}

\def\aj{{AJ}}

\def\nat{{Nature}}
\def\pasp{{PASP}}

\def\mnras{{MNRAS}}

\def\aap{{A\&A}}

\def\eps@scaling{.95}
\def\epsscale#1{\gdef\eps@scaling{#1}}

\def\plotone#1{\centering \leavevmode
    \epsfxsize=\eps@scaling\columnwidth \epsfbox{#1}}

\def\plotfiddle#1#2#3#4#5#6#7{\centering \leavevmode
    \vbox to#2{\rule{0pt}{#2}}
    \includegraphics{#1}}

\begin{document}
\pagenumbering{arabic}

\title{High-Redshift, Radio-Loud Quasars}

\titlerunning{High-Redshift, Radio-Loud Quasars}

\author{Daniel Stern\inst{1,2}}

\institute{Department of Astronomy, University of California at
Berkeley, 601 Campbell Hall, Berkeley CA~94720
\and
Jet Propulsion Laboratory, California Institute of Technology, MS
169-327, Pasadena CA~91109
[{\tt e-mail:  stern@zwolfkinder.jpl.nasa.gov}]}

\authorrunning{Stern }

\maketitle

\begin{abstract}

I discuss two programs to study radio-loud quasars at high ($z \simgt
4$) redshift.   Quasars are the most luminous, non-transient objects
known and are observed to the earliest cosmic epochs.  At lower
redshifts, radio-loud quasars are associated with host galaxies having
de~Vaucoleurs profiles.  By association, identifying and studying a
sample of high-redshift, radio-loud quasars provides important clues to
the early Universe and potentially probes early-type galaxy formation.
The first aspect of this proceeding discusses an extensive search for
high-redshift, radio-loud quasars culled from matching the FIRST radio
catalog with the 2$^{nd}$ generation Palomar Sky Survey.  The second
aspect of this proceeding concerns studying the radio properties of
optically-selected $z > 4$ quasars.  At the time of this conference,
only seven $z > 4$ quasars were in the literature.  The first program
discussed herein uncovers three new such sources, while the second
program identifies an additional five.  We use the new samples to study
the evolution of the radio-loud fraction, finding no evolution in this
quantity between redshift $z \simeq 2$ and $z > 4$ for $-26 > M_B >
-28$.

\end{abstract}

\section{Introduction}

High-redshift quasars provide some of the earliest glimpses we have of the
Universe, constrain models of structure formation, and are valuable probes
of the intervening intergalactic medium (IGM).  Quasars were originally
identified from their radio emission (Schmidt 1963).  Subsequent work
revealed that only $\sim$ 10\%\, of quasars are radio-loud, resulting
in a de-emphasis of radio selection in quasar searches.  Most known
quasars have been discovered solely from optical observations.  Possible
obscuration and absorption processes in quasars, however, imply that
radio is perhaps a better search wavelength for quasars, less prone to
selection effects than optical surveys.  Radio selection has resulted
in the discovery of objects of unusual spectral character such as the
first radio-loud broad absorption line quasar, FIRST~J155633+351758
(Becker {et~al.} 1997).  Webster {et~al.} (1994) suggests that a large population
of dust-reddened quasars exists and has escaped detection in optical
(grism) surveys.  This discovery, if confirmed, would have serious
implications for dust formation in the early Universe and provide an
opportunity to learn about both the quasar environment and the degree
to which these objects contribute to the X-ray and gamma-ray backgrounds.

\section{A Search for $z \approx 4$ Radio-Loud Quasars}

I first report\footnote{This section reports on work in progress being
pursued in collaboration with Steve Odewahn, Roy Gal, S.G. Djorgovski,
R.  de~Calvalho, Wil van~Breugel, \& Hyron Spinrad.  Details of this
program will be presented in Stern et~al. (2000b).} on a systematic
search for high-redshift, radio-loud quasars selected using the 1.4 GHz
VLA FIRST survey (Faint Images of the Radio Sky at Twenty-one
Centimeters; Becker, White, \& Helfand 1995) in conjunction with the
(second generation) digitized Palomar Observatory Sky Survey (DPOSS;
Djorgovski {et~al.} 1996).  The purpose of this program is to augment
the census of such sources, provide a sample for follow-up studies, and
study evolution in the radio-loud quasar population.  Considering
optical counterparts to FIRST radio sources, we spectroscopically
target unresolved optical identifications with colors consistent with
distant quasars.  Previous studies have generally focussed on
flat-spectrum sources with flux densities $1 - 3$ orders of magnitude
greater than the FIRST survey limit and have unveiled seven radio-loud
quasars at $z > 4$ (Schneider {et~al.} 1992; McMahon {et~al.} 1994;
Hook {et~al.} 1995; Shaver, Wall, \&  Kellerman 1996a; Zickgraf
{et~al.} 1997; Hook \& McMahon 1998).  This work has suggested a
decrease in the space density of radio-loud quasars at high redshift
(\eg Shaver {et~al.} 1999, but see Rawlings, this volume).  Our sample
probes a new regime of parameter space:  low radio flux density,
high-redshift quasars irregardless of radio spectral index.  This last
point is by necessity since no large-area radio survey of comparable
depth exists at a frequency different from FIRST.  This work helps
constrain the radio-faint end of the radio quasar luminosity function
at high redshift, and provides a useful sample of and to the distant
Universe.

\begin{figure}[!h]
\plotone{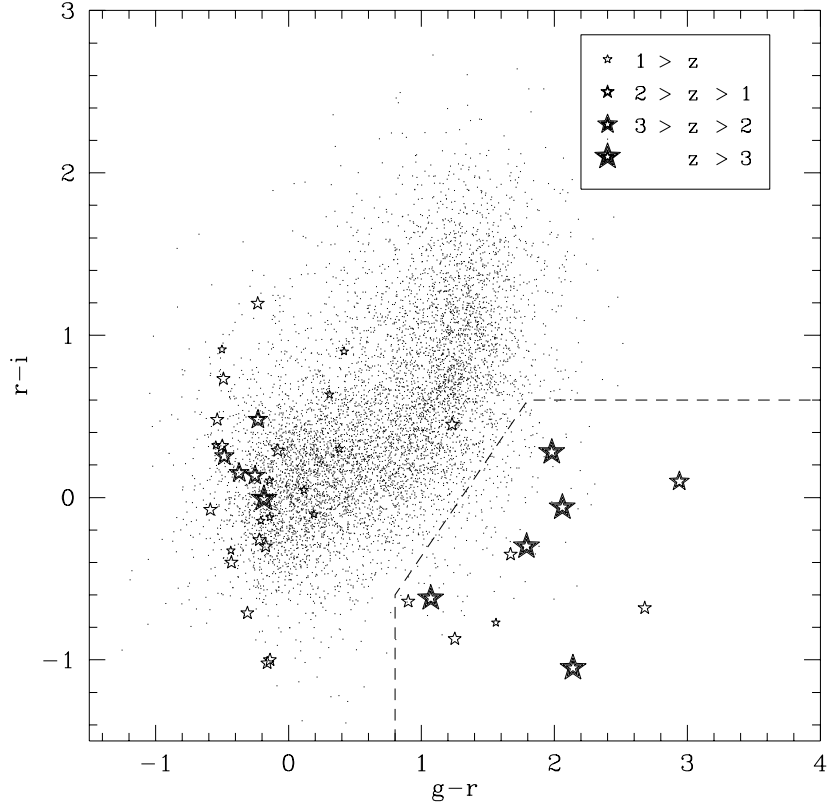}

\caption{Color-color diagram for objects classified as stellar by
DPOSS.  Quasars, indicated by stars with the symbol size proportional
to the redshift, include sources both from the current program and the
literature.  We select objects in the bottom right of the diagram
identified with FIRST radio sources for spectroscopic follow-up.}

\label{sternqso1}
\end{figure}

\subsection{Target Selection}

Our work relies on two important, large-area sky surveys of unprecedented
accuracy, sky-coverage, and depth.  Both surveys are still in production
mode and the current work has been has been made possible by the generous
collaboration of the principal investigators and research associates of
both surveys.

The FIRST survey (Becker {et~al.} 1995) is a Very Large Array (VLA) effort
to map the northern sky with 1\farcs0 positional accuracy to a limiting
flux of $S_{\rm 1.4GHz} = 1$ mJy (5~$\sigma$).  The radio observations,
obtained with the B-array of the VLA, have a resolution of 4\farcs0
and, as of 1999 July, the catalog contains $\sim 550,000$ sources over
$\sim 6000$ square degrees.

The Second Palomar Observatory Sky Survey (POSS-II, or DPOSS for
the digitized version; Djorgovski {et~al.} 1999) is a photographic montage of
the northern sky obtained with the 48\arcsec\ Schmidt telescope at
Palomar Observatory.  The survey encompasses three pass bands:  blue
($J$; $\lambda_{\rm eff} \sim 4800$ \AA), red ($F$; $\lambda_{\rm eff}
\sim 6500$ \AA), and infrared ($N$; $\lambda_{\rm eff} \sim 8500$ \AA),
which are transformed to standard Gunn $gri$ magnitudes with
calibration CCD images obtained at the Palomar Observatory
60\arcsec\ telescope.  Typical limiting magnitudes are 22.5, 20.8, and
19.5 respectively ($3\sigma$) -- \ie POSS-II samples $1 - 1.5$
magnitudes deeper than the POSS-I and includes an additional redder
plate.  This depth is comparable to that probed by the Sloan Digital
Sky Survey, but contains two fewer pass bands.  The calibration CCD
images are also used to train a neural net to distinguish stars and
galaxies (\eg Odewahn {et~al.} 1996).   

Optical identifications were made based on positional coincidences
of the FIRST radio position with an optical counterpart on the DPOSS
$r$-band plate.  For each 6\deg\ $\times$ 6\deg\ DPOSS plate, there
are approximately 3800 FIRST radio sources, of which $\approx$ 20\%\
have optical identifications.  Our criterion for positional coincidence
is $\Delta r \leq 3\farcs0$.  For each 36 square degree DPOSS plate, our
selection criterion yields $\approx 800 \pm 100$ optical identifications
of FIRST sources.

Fig.~\ref{sternqso1} presents the color-color diagram for objects
classified as stellar by DPOSS in the $r$-band, with confirmed quasars
both from our work and the literature indicated.  A broad stellar locus
is clearly evident.  High-redshift quasars are selected based on red $g -
r$ color, attributable to absorption from the intervening \lya\ forest
at $3.8 \simlt z \simlt 4.6$.  We also demand sources be relatively
flat in $r - i$ to minimize contamination from Galactic M-stars.
These criteria are relatively robust, and similar to the criteria used
by Kennefick, Djorgovski, \&  de~Calvalho (1995); Kennefick, Djorgovski, \&  Meylan (1996) in their multicolor program to
find distant radio-quiet quasars.

We find approximately $1 - 2$ identifications per DPOSS plate that
meet our selection criteria, namely:  (i) $S_{\rm 1.4GHz} > 1$ mJy,
(ii) positional coincidence, $\Delta r \leq 3\farcs0$, (iii) unresolved
on $r$-band DPOSS plate, (iv) $r > 17$, and (v) red optical color.

\begin{figure}[!h]
\plotone{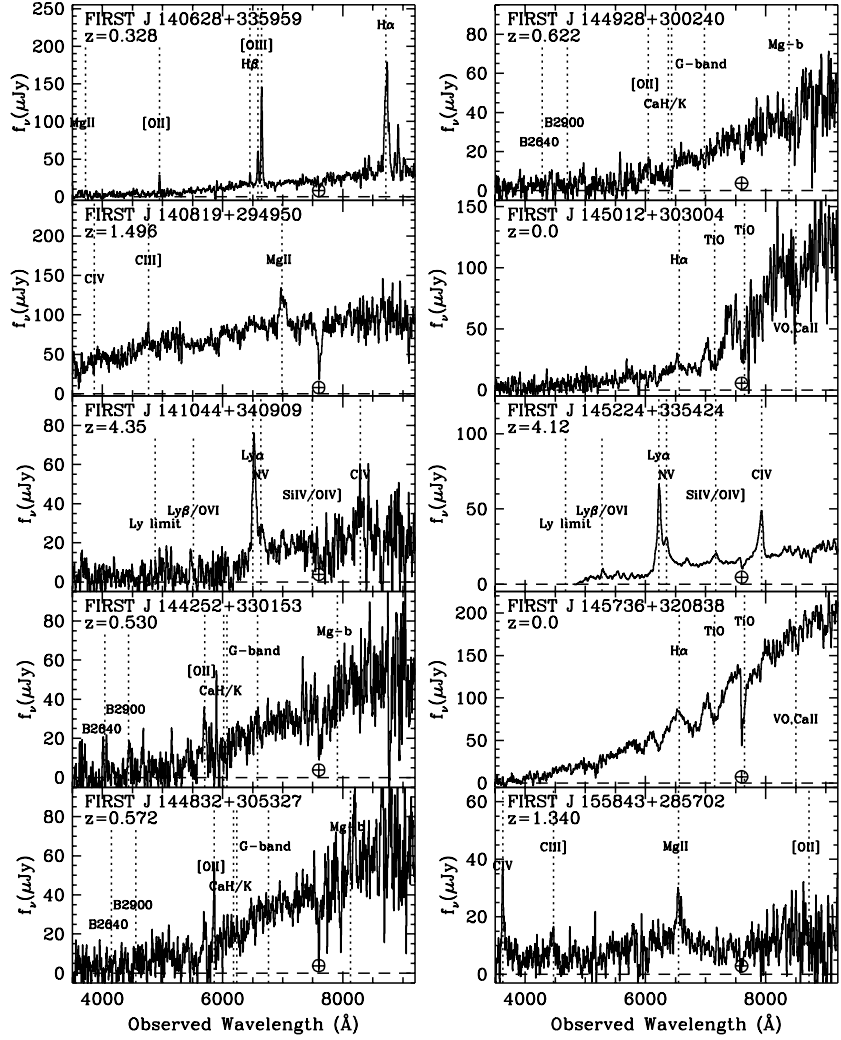}

\caption{Results of a ``typical'' observing run (UT 1998 June 26).
Note the range in objects identified:  late-type Galactic stars,
early-type galaxies, strong line-emitting galaxies, and quasars.  The
early-type galaxies are particularly difficult to identify at this
signal-to-noise ratio; their redshifts should be considered tentative.
The crossed circle indicates the wavelength of the telluric A-band.}

\label{sternqso2}
\end{figure}

\subsection{Results and Discussion}

We have obtained spectra of 153 FIRST/DPOSS targets.  Most
spectroscopic observations of candidates were carried out in the period
$1996 - 1999$ using the Kast imaging double-spectrograph
(Miller \& Stone 1994) at the Cassegrain focus of the 3m Shane Telescope at
Lick Observatory.  Fig.~\ref{sternqso2} presents the results from a
``typical'' observing run.  A handful of sources were also observed at
the 5m Hale Telescope at Palomar Observatory.  and with the Low
Resolution Imaging Spectrometer (LRIS; Oke {et~al.} 1995) at Keck
Observatory.

Six new high-redshift ($z > 3.8$) quasars have been discovered,
including three at $z > 4$.  The remaining targets include 16
lower-redshift quasars, $\approx 60$ moderate-redshift ($z \sim 0.5$)
evolved galaxies for which the 4000 \AA\ break mimics a high-redshift
\lya\ forest, 29 Galactic M stars, 11 emission line galaxies, and 33
unidentified spectra.  The last objects are apt to be evolved galaxies
lacking strong emission features which are difficult to identify
reliably at low signal-to-noise ratio, though the possibility of a
high-redshift quasar, or blazar cannot be ruled out.  During inclement
conditions, we observed brighter FIRST/DPOSS targets without bias with
respect to color.  Many of the lower-redshift quasars derive from that
sample.

\begin{figure}[!ht]
\plotone{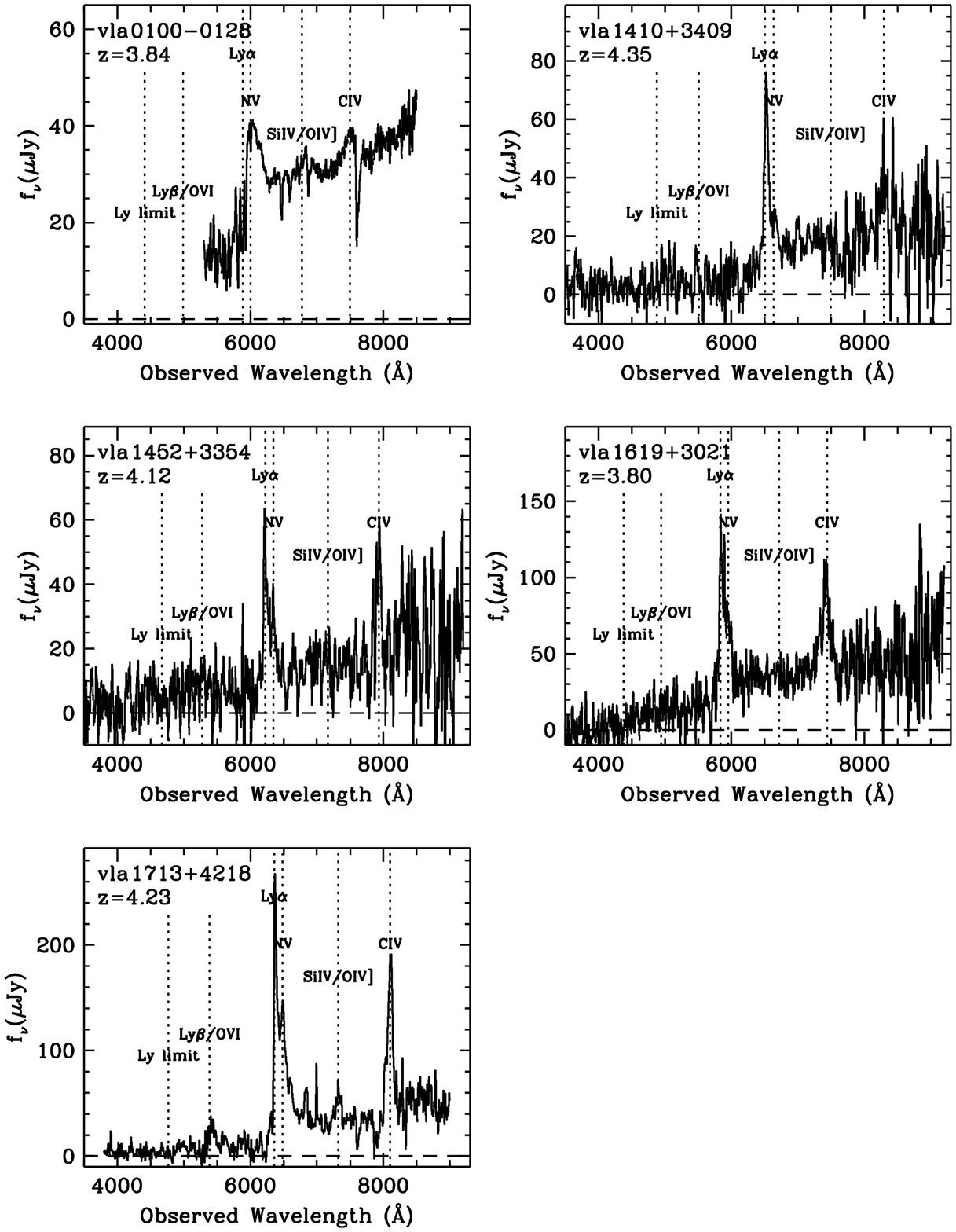}

\caption{Spectra of the new high-redshift quasars.}

\label{sternqso3}
\end{figure}

%

\begin{figure}[!h]
\plotfiddle{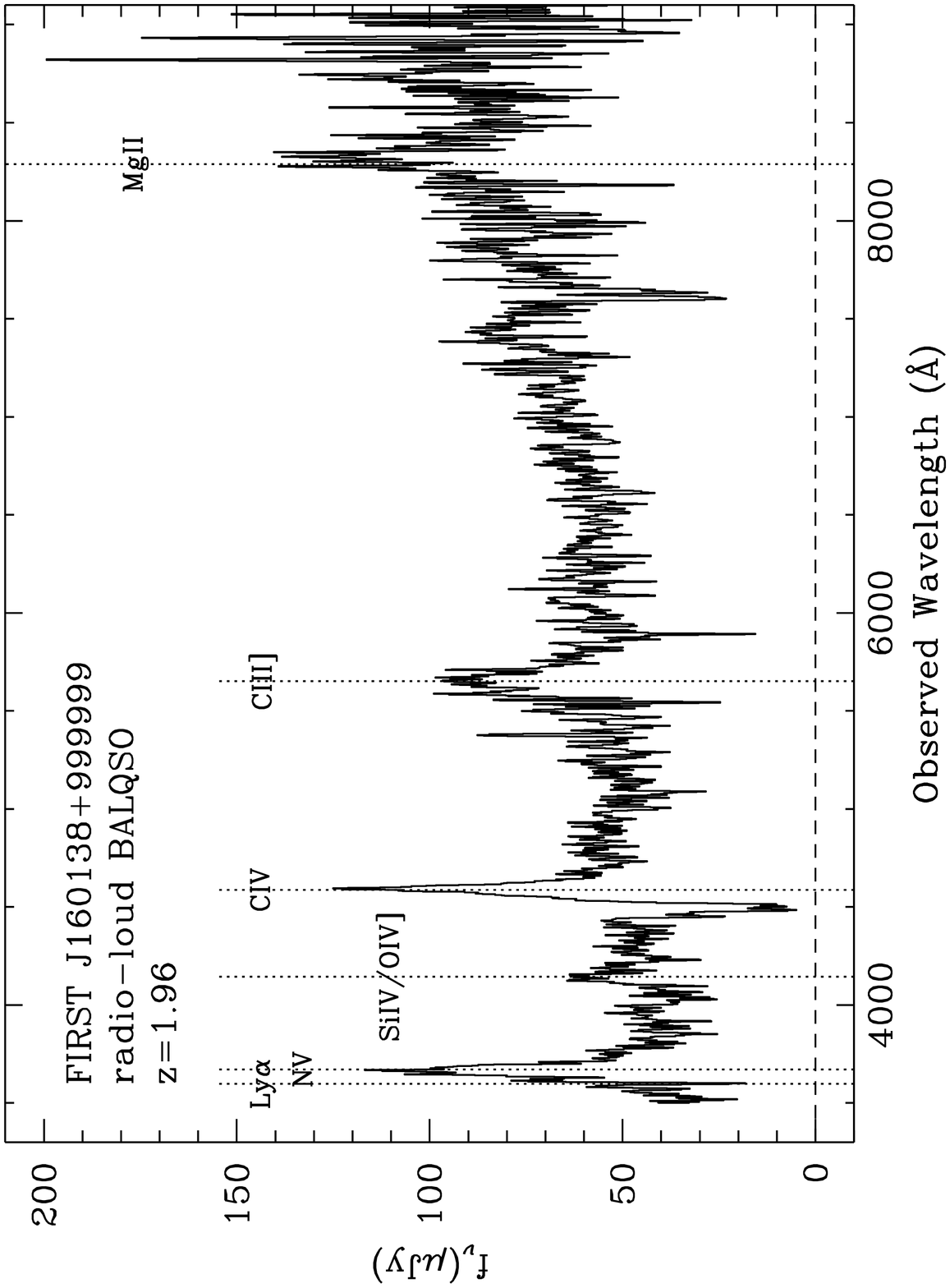}{3.2in}{-90}{40}{40}{-160}{255}

\caption{Spectrum of the radio-loud broad absorption line quasar
(BALQSO) FIRST~J160138+294734, identified from this survey.}

\label{sternqso4}
\end{figure}

The spectra of the six new high-redshift quasars are shown in
Fig.~\ref{sternqso3}.  Fig.~\ref{sternqso4} presents a spectrum of
FIRST~J160138+294734, one of the more interesting lower-redshift sources
uncovered from this survey.  It is an example of a radio-loud, broad
absorption line quasar (BALQSO), a class of object whose existence was
questioned until 1997.  The implications of this study on the radio-loud
quasar luminosity function is discussed in Stern {et~al.} (2000b).

\section{Radio Properties of $z>4$ Optically-Selected Quasars}

The second aspect\footnote{This section is a synopsis of Stern
{et~al.} (2000a), and presents work done in collaboration with S.G.
Djorgovski, Rick Perley, Reinaldo de~Calvalho, \& J.V. Wall.} of this
proceeding discusses statistical studies of the radio properties of $z
> 4$ optically-selected quasars.  Already in the late 1960s it was
becoming apparent that the rapid increase in quasar comoving space
densities with redshift did not continue beyond $z \sim 2$.  Several
authors suggested the existence of a redshift cutoff beyond which a
real decrease in quasar densities occurs, unrelated to observational
selection effects (\eg Sandage 1972).  Numerous studies have now shown
that this is indeed the case.  For example, Schmidt, Schneider, \&
Gunn (1995a), applying the $V/V_{\rm max}$ test to a well-defined
sample of 90 quasars detected by their \lya\ emission in the Palomar
Transit Grism Survey, conclude that the comoving quasar density
decreases from $z = 2.75$ to $z = 4.75$.  Shaver {et~al.} (1996b) show
a similar decrease in space density exists for radio-loud quasars,
implying that the decline is not simply due to obscuration by
intervening galaxies.

An unresolved question, however, is how the decline compares between
the optically-selected and radio-selected populations.  Differential
evolution between the radio-quiet and radio-loud quasar populations
would be a fascinating result, which could change (or challenge) our
understanding of different types of active galactic nuclei (AGN).  For
example, Wilson \& Colbert (1995) propose a model whereby radio-loud AGN are
products of coalescing supermassive black holes in galaxy mergers; such
a process would result in rapidly spinning black holes, capable of
generating highly-collimated jets and powerful radio sources.  Such a
model could naturally explain a lag (if there is indeed one) between
the appearance of powerful radio-loud quasars in the early Universe and
the more common radio-quiet quasars.

Several groups have reported on follow-up of optically-selected quasar
surveys at radio frequencies (Kellerman {et~al.} 1989; Miller, Peacock, \& Mead 1990; Schneider {et~al.} 1992; Visnovsky {et~al.} 1992; Hooper {et~al.} 1995; Schmidt {et~al.} 1995b; Goldschmidt {et~al.} 1999).
Typically between 10\%\ and 40\%\ of the quasars are detected in the
radio, but cosmological evolution in this quantity remains uncertain.
Some researchers find evidence that the radio-loud fraction of
optically-selected quasars decreases with increasing redshift
(\eg Visnovsky {et~al.} 1992), while others find no (or only marginal)
evidence for evolution (\eg {La~Franca} {et~al.} 1994; Goldschmidt {et~al.} 1999), and
yet other researchers find that the radio-loud fraction {\em increases}
with increasing redshift (\eg Hooper {et~al.} 1995).  We attempt to
re-address the issue by providing a more accurate measure of the
radio-loud fraction at $z > 4$.

\subsection{Targeted 4.8 GHz VLA Mapping}

To determine the radio-loud fraction of $z > 4$ quasars, we observed 32
high-redshift quasars at 4.8~GHz with the VLA.  The quasar sample is
from the Palomar multicolor survey (Kennefick {et~al.} 1995; Djorgovski {et~al.} 1999)
and represents all $z > 4$ quasars found from that survey at the time
of the radio observations.  Six of the quasars are equatorial objects
found previously in a similar search by the Automated Plate Machine
(APM) group (Irwin, McMahon, \& Hazard 1991) which fall within the Palomar multicolor
survey selection criteria.  These 32 sources comprise a statistically
complete and well-understood sample of optically-selected $z > 4$
quasars;  they should be an unbiased set as far as the radio properties
are concerned.

The VLA observations were made on UT 1997 March 14 at 4835 and 4885~MHz.
Calibration and imaging followed standard procedures.  Each target was
observed for $30 - 45$~min, including calibration, yielding typical rms
noise between 30 and 60 $\mu$Jy.

We detect four quasars from this sample, where a match is liberally
defined to correspond to a radio source lying within 10\arcsec\ of the
optical source.  In fact, the four matches all have optical-to-radio
positional differences less than 2\arcsec, implying that the
identifications are unlikely to be spurious.  Three sources represent
newly identified high-redshift, radio-loud quasars.

\subsection{High-Redshift Quasars in FIRST and NVSS}

To augment the targeted VLA sample discussed above, we also correlated an
updated list of all 134 $z > 4$ quasars known to us in mid-1999 with two
recent, large-area radio surveys:  the FIRST survey (Becker {et~al.} 1995),
discussed in \S 2.1, and the 1.4~GHz NVSS survey (Condon {et~al.} 1998),
which covers a larger area to a shallower flux density limit ($S_{\rm
1.4 GHz}^{\rm lim} = 2.25~{\rm mJy}, 5\sigma$).  For luminosity distance
$d_L$ and radio spectral index $\alpha$ ($S_\nu \propto \nu^\alpha$),
the restframe specific luminosity $L_\nu = 4 \pi d_L^2 S_\nu / (1 +
z)^{1 + \alpha}$, where both $L_\nu$ and $S_\nu$ are measured at the
same frequency.  At $z = 4$ and for a typical quasar spectral index of
$\alpha = -0.5$, the FIRST survey reaches a 3$\sigma$ limiting 1.4~GHz
specific luminosity of $\log L_{1.4}~ (h_{50}^{-2} \ergsHz) = 32.6$.
The comparable limit for the NVSS survey is $\log L_{1.4}~ (h_{50}^{-2}
\ergsHz) = 32.9$.  Therefore, using the radio luminosity definition of
radio loudness, $L_{\rm 1.4 GHz} \geq 10^{32.5}~ h_{50}^{-2}$ \ergsHz
(Gregg {et~al.} 1996), these surveys incompletely sample radio-loud quasars
at high redshift.

\begin{table}[ht!]
\caption{High-Redshift Quasars Detected by FIRST}
\footnotesize
\begin{center}
\begin{tabular}{lcccccc}
\hline\hline
&
&
$r$ &
$M_B$ &
$S_{\rm 1.4 GHz}$ &
Offset &
\\
Quasar &
$z$ &
(mag) &
(mag) &
(mJy) &
(arcsec) &
Ref \\
\hline
BRI0151$-$0025&4.20 & 18.9 & $-27.55$ & $4.75 \pm 0.15$ & 0.90 & 1$-\star$ \\
PSS1057+4555 & 4.12 & 17.7 & $-28.65$ & $1.38 \pm 0.13$ & 2.13 & 2$-\star$ \\
FIRST~J141045+34009 & 4.35 & 19.6 & $-27.10$ & $2.07 \pm 0.15$ & 0.84 & 3 \\
GB1428+4217  & 4.72 & 20.9 & $-26.44$ & $215.62 \pm 0.15$&0.41 & 4 \\
FIRST~J145224+335424 & 4.12 & 20.4 & $-25.95$ & $6.95 \pm 0.13$ & 3.02 & 3 \\
FIRST~J171356+421808 & 4.23 & 19.0 & $-27.49$ & $2.80 \pm 0.15$ & 0.00 & 3 \\
\hline
\end{tabular}
\end{center}
\medskip

\emph{Notes.---}  Newly identified high-redshift, radio-loud quasars
are indicated with a $\star$.  Positions are from the FIRST catalog.
References:  (1) Smith {et~al.} (1994);  (2) Kennefick {et~al.} (1995); (3)
Stern {et~al.} (2000b); (4) Hook \& McMahon (1998).

\end{table}
\normalsize

Of the 134 $z > 4$ quasars, 51 reside in portions of the celestial
sphere observed thus far by FIRST.  We deemed a radio detection in
FIRST to be associated with a high-redshift quasar if it lay within
10\arcsec\ of the quasar optical position.  Six high-redshift, radio-loud
quasars were identified, of which four were previously known (Table 1).
One source, PSS~1057+4555, was {\em undetected} in the targeted 5~GHz
survey, implying an unusually steep spectral index, $\alpha < -2.05$
(3 $\sigma$) or significant variability on the time scale corresponding
to the epochs of the two radio surveys.

\begin{table}[ht!]
\caption{Optically-Selected, High-Redshift Quasars Detected by NVSS}
\footnotesize
\begin{center}
\begin{tabular}{lcccccc}
\hline\hline
&
&
$r$ &
$M_B$ &
$S_{\rm 1.4 GHz}$ &
Offset &
\\
Quasar &
$z$ &
(mag) &
(mag) &
(mJy) &
(arcsec) &
Ref \\
\hline
PC0027+0525  & 4.10 & 21.5 & $-24.82$ & $4.8 \pm 0.5$ & 8.47 & 1$-\star$ \\
PSS0121+0347 & 4.13 & 17.9 & $-28.46$ & $78.6 \pm 2.4$& 0.76 & 2$-\star$ \\
BRI0151$-$0025&4.20 & 18.9 & $-27.55$ & $3.0 \pm 0.5$ & 7.57 & 3$-\star$ \\
RXJ1028.6$-$0844&4.28&18.9 & $-27.67$ & $269.8 \pm 8.1$ & 22.4 & 4 \\
BRI1050$-$0000&4.29 & 18.8 & $-27.78$ & $9.7 \pm 0.6$ & 3.04 & 5,6 \\
BRI1302$-$1404&4.04 & 18.6 & $-27.67$ & $20.6 \pm 0.8$  & 0.61 & 7$-\star$ \\
FIRST~J141045+340909 & 4.36 & 19.6 & $-27.10$ & $3.5 \pm 0.15$ & 13.5 & 8 \\
GB1428+4217  & 4.72 & 20.9 & $-26.44$ & $211.1 \pm 6.3$ & 1.10 & 9 \\
FIRST~J145224+335424 & 4.12 & 20.4 & $-25.95$ & $8.6 \pm 0.5$ & 2.42 & 8 \\
GB1508+5714  & 4.30 & 18.9 & $-27.70$ & $202.4 \pm 6.1$ & 22.3 & 10 \\
FIRST~J171356+421808 & 4.23 & 19.0 & $-27.49$ & $2.9 \pm 0.5$ & 3.40 & 8 \\
GB1713+2148  & 4.01 & 21.0 & $-25.25$ & $447.2 \pm 15.0$ & 2.02 & 9 \\
RXJ1759.4+6638&4.32 & 19.0 & $-27.63$ & $3.6 \pm 0.7$ & 11.6 & 11$-\star$ \\
PC2331+0216  & 4.10 & 20.0 & $-26.32$ & $2.7 \pm 0.5$ & 6.51 & 12,13 \\
\hline
\end{tabular}
\end{center}
\medskip

\emph{Notes.---}  Newly identified high-redshift, radio-loud quasars
are indicated with a $\star$.  Positions are from the NVSS catalog.
References:  (1) Schneider, Schmidt, \& Gunn (1997); (2) Djorgovski {et~al.} (1999); (3)
Smith {et~al.} (1994); (4) Zickgraf {et~al.} (1997); (5) Smith {et~al.} (1994); (6)
McMahon {et~al.} (1994); (7) Storrie-Lombardi {et~al.} (1996); (8)
Stern {et~al.} (2000b); (9) Hook \& McMahon (1998); (10) Hook {et~al.} (1995); (11)
Henry {et~al.} (1994); (12) Schneider, Schmidt, \& Gunn (1989); (13)
Schneider {et~al.} (1992).  Optical coordinates for BRI1050$-$0000 are from
Storrie-Lombardi {et~al.} (1996), which are $\approx$ 50\arcsec\ offset from
the coordinates reported in the discovery paper (Smith {et~al.} 1994).

\end{table}
\normalsize

Of the 134 $z > 4$ quasars, 129 reside in portions of the celestial
sphere observed thus far by NVSS.  A radio detection in NVSS was deemed
associated with a high-redshift quasar if it lay within 30\arcsec\ of the
quasar optical position.  We found a total of 14 radio identifications
(Table 2); 5 of these represent newly-identified high-redshift,
radio-loud quasars.

\begin{figure}[!h]
\plotfiddle{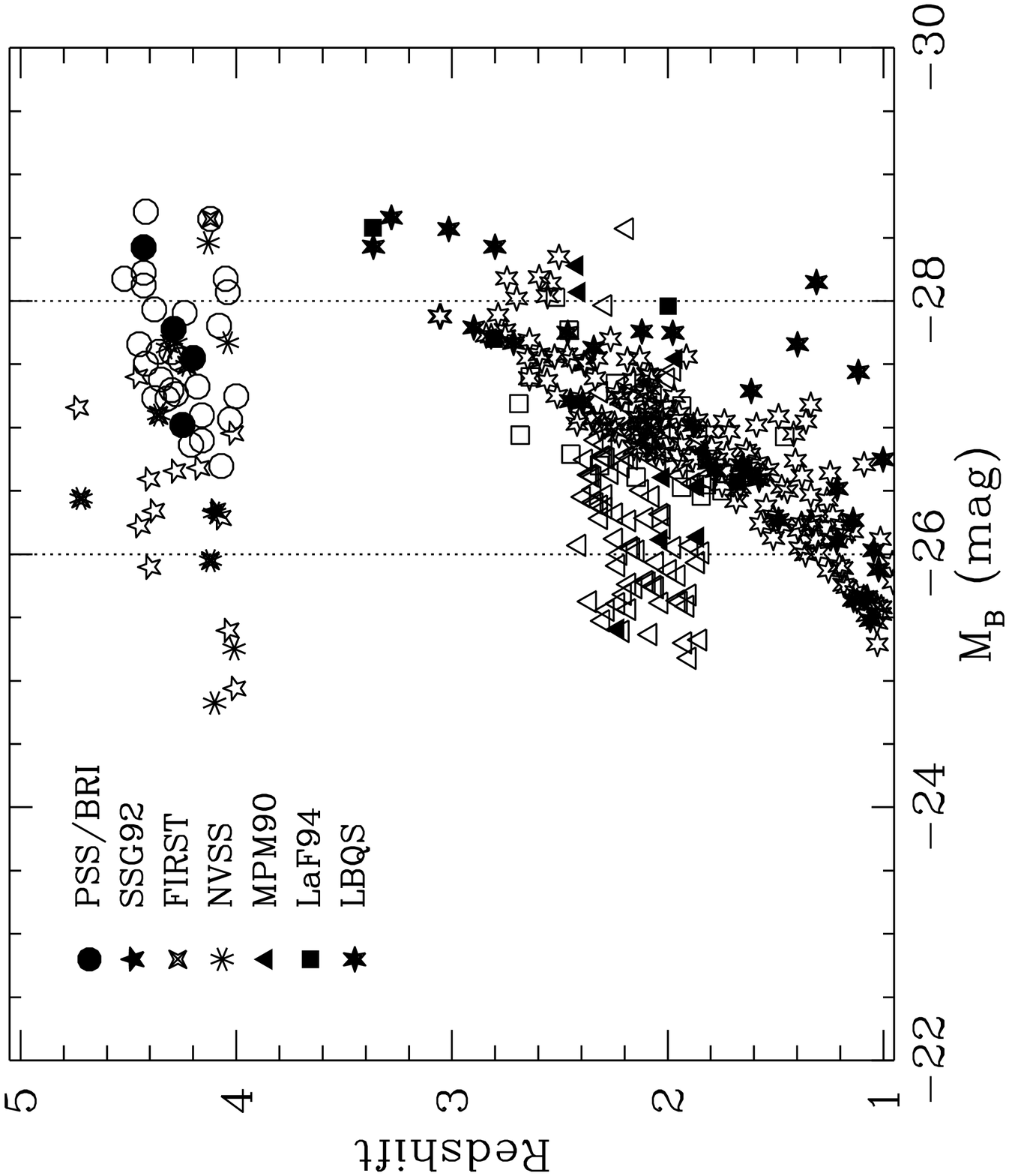}{3.2in}{-90}{40}{40}{-160}{255}

\caption{Location of all quasars considered here in the optical
luminosity--redshift plane.  PSS/BRI refers to the 5~GHz deep imaging
discussed in \S 3.1.  See text for description of other surveys.
Non-detections are indicated with open symbols.  The $z > 4$ FIRST and
NVSS non-detections are not presented.  Vertical dotted lines indicate
the optical luminosity range considered in our analysis.}

\label{sternqso5}
\end{figure}

\subsection{Results}

Fig.~\ref{sternqso5} summarizes the location of the $z > 4$ quasars in
the optical luminosity--redshift plane.  Consonant with their extreme
distances, these quasars have extremely bright absolute magnitudes,
$M_B \simlt -26$.  To probe evolution in the radio-loud quasar
fraction, we require comparison with a sample of comparably luminous
quasars at lower redshift.  Fig.~\ref{sternqso5} illustrates such a
sample:  we have augmented the $z > 4$ quasars with several
lower-redshift surveys from the literature:  SSG92 refers to the
Schneider {et~al.} (1992) sample,  MPM90 refers to Miller {et~al.} (1990),  LaF94
refers to {La~Franca} {et~al.} (1994), and  LBQS refers to the 8.4~GHz VLA imaging
of LBQS quasars reported in Visnovsky {et~al.} (1992) and
Hooper {et~al.} (1995).  Luminosities have been calculated assuming that
both the radio spectral index $\alpha$ and the optical spectral index
$\alpha_{\rm opt}$ are equal to $-0.5$ (for details,
see Stern {et~al.} 2000a).

\begin{figure}[!h]
\plotfiddle{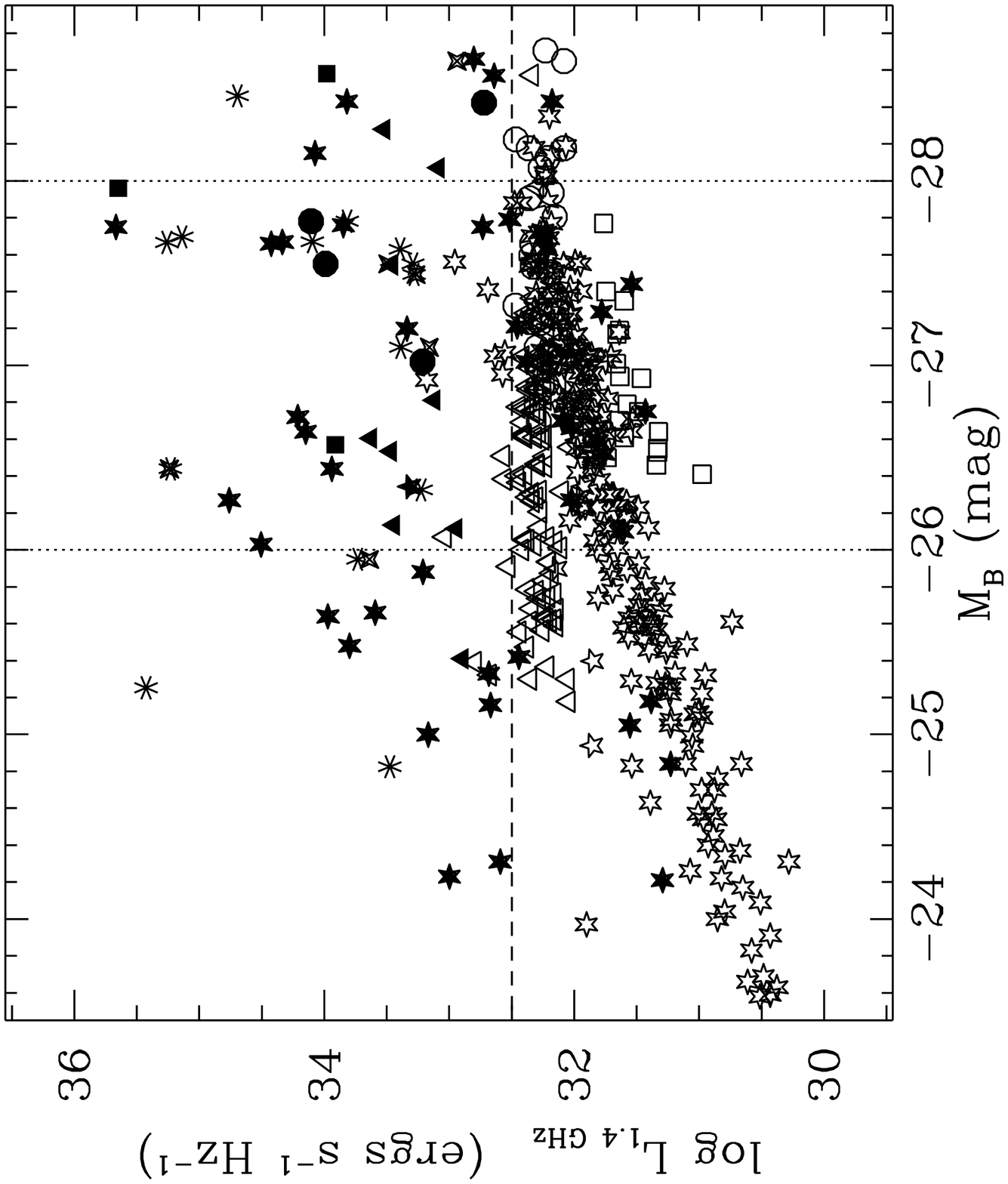}{3.2in}{-90}{40}{40}{-160}{255}

\caption{Location of all quasars considered here in the optical
luminosity--radio luminosity plane.  Symbols are as in
Fig.~5; again, non-detections are indicated with
open symbols.  Horizontal dashed line refers to the cutoff between
radio-loud and radio-quiet quasars.  A few sources with elevated noise
levels are classified as radio-loud.  Vertical dotted lines indicate
the optical luminosity range considered in our analysis.  }

\label{sternqso6}
\end{figure}

Fig.~\ref{sternqso6} illustrates the location of the quasars in
$M_B - L_{\rm 1.4 GHz}$ space.  For the purposes of the following
analysis, we assume non-detections have radio fluxes equal to their $3
\sigma$ noise value.  We omit the new $z > 4$ radio-loud quasars
detected by FIRST and NVSS in the following analysis, as those surveys
are insufficiently sensitive to reach the radio-loud cutoff at $z >
4$.  The detection limits of the other surveys are sufficiently deep
that few non-detections are above the radio-loud/radio-quiet boundary;
we conservatively classify those few sources as radio-loud below.  Our total
sample is 428 quasars spanning $0.2 < z < 4.7$, $-22.7 > M_B > -28.7$,
and $30.08 < \log L_{\rm 1.4 GHz} (h_{50}^{-2} \ergsHz) < 35.7$.

\begin{figure}[!h]
\plotfiddle{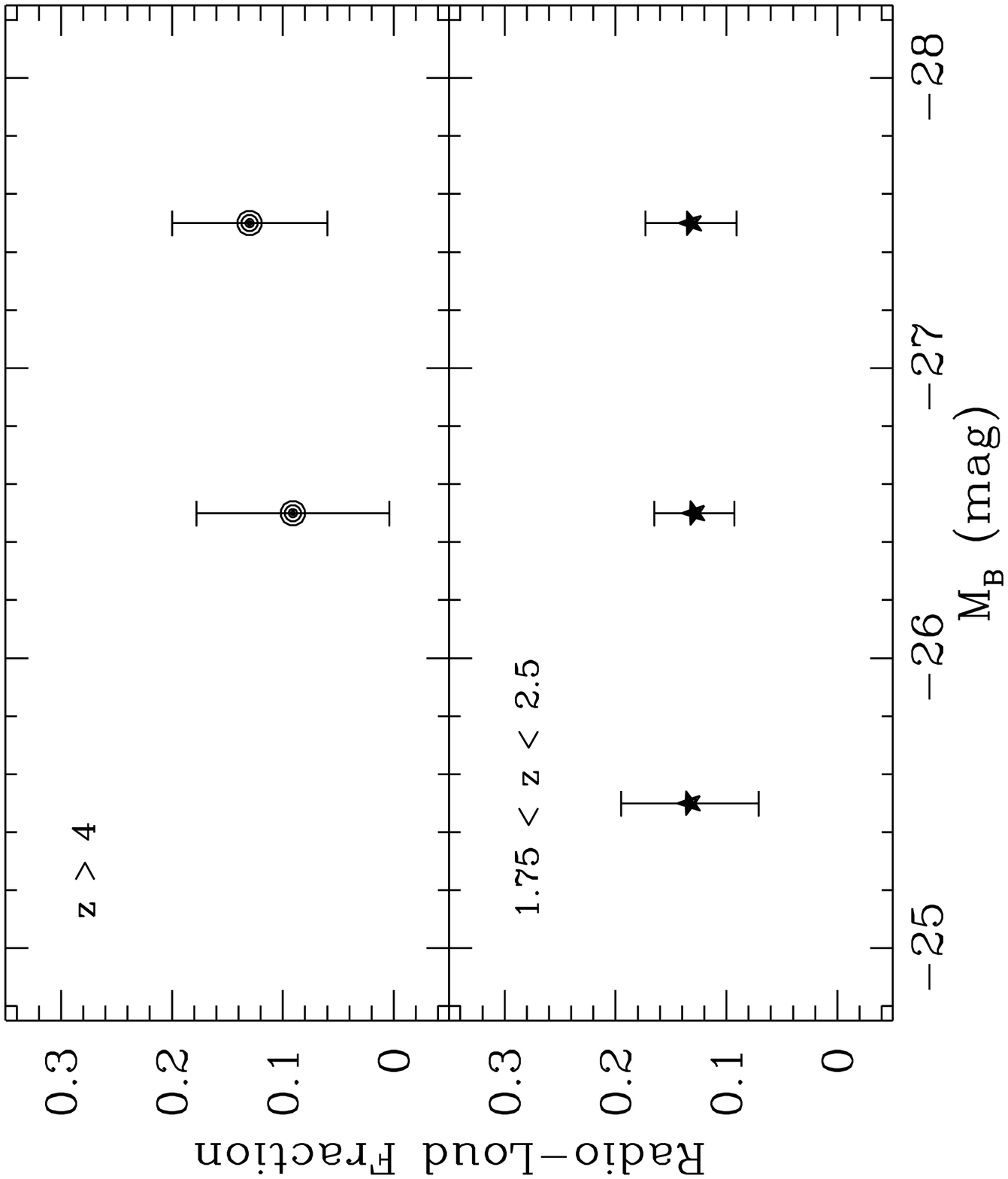}{3.2in}{-90}{40}{40}{-160}{255}

\caption{Fraction of radio-loud quasars as function of absolute
magnitude.  Two redshift ranges are plotted with 1~mag wide bins.  No
evidence for luminosity dependence in the radio-loud fraction is
evident from this limited data set.}

\label{sternqso7}
\end{figure}

We first consider if we detect optical luminosity dependence in the
radio-loud fraction.  We consider two redshift ranges where we have
large samples of quasars, and divide the sample into absolute magnitude
bins where the quasars are {\em approximately} smoothly distributed.
Fig.~\ref{sternqso7} shows this fraction for each redshift bin.  Error
bars shown are the square root of the variance, $f (1-f) / N$, where
$f$ is the radio-loud fraction and $N$ is the number of quasars in the
bin considered (\eg Schneider {et~al.} 1992).  The impression from
Fig.~\ref{sternqso7} is that for a given redshift bin, the radio-loud
fraction is independent of optical luminosity.  This result stands in
contrast to the analysis of Goldschmidt {et~al.} (1999) who find that the
radio-loud fraction increases with luminosity for each redshift bin
considered.  Consideration of the radio-loud fraction plotted in Fig.~7
of their paper suggests that the optical luminosity dependence claimed
at $1.3 < z < 2.5$ depends largely on the poorly measured radio-loud
fraction at $M_B = -28$.  However, at $0.3 < z < 1.3$, their data
convincingly shows optical luminosity dependence in the radio-loud
fraction.  Comparing the ordinate between the two panels of
Fig.~\ref{sternqso7} below suggests that the radio-loud fraction remains
approximately constant with redshift; we consider this next.

\begin{figure}[!h]
\plotfiddle{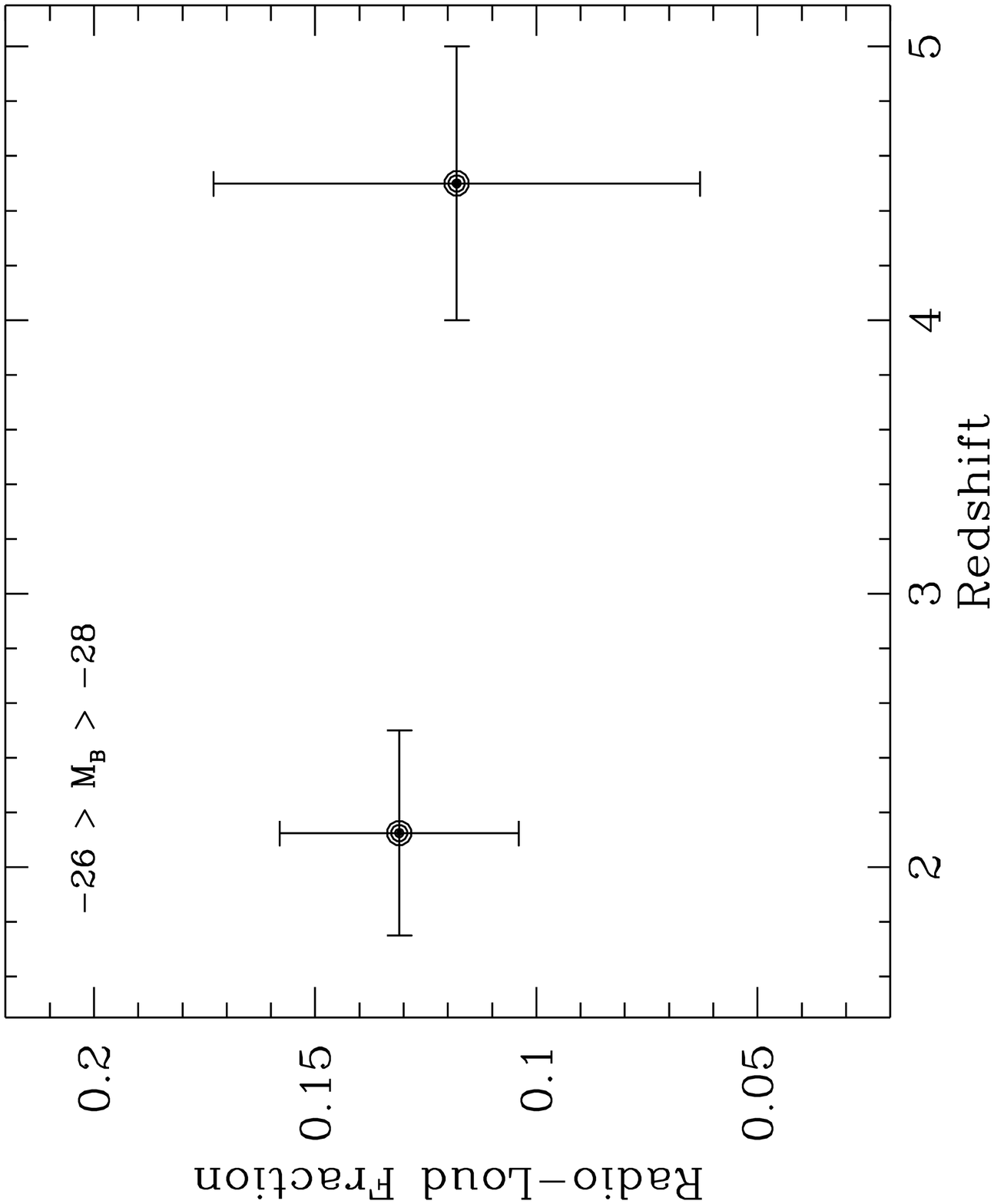}{3.2in}{-90}{40}{40}{-160}{255}

\caption{Fraction of radio-loud quasars as function of redshift.  Two
redshift ranges are plotted.  No redshift dependence in the radio-loud
fraction is evident.}

\label{sternqso8}
\end{figure}

Since there is little evidence of optical luminosity dependence of the
radio-loud fraction in our data set, we decrease our errors by
considering the radio-loud fraction in a larger absolute magnitude
range, $-26 < M_B < -28$.  Fig.~\ref{sternqso8} shows the results of
this analysis, with error bars calculated as above.  At $1.75 < z <
2.5$, we classify 20 out of the 153 quasars in the optical luminosity
range considered as radio-loud, corresponding to $13.1 \pm 2.7$\%\ of
the quasars.  At $z > 4$, we classify 4 out of the 34
optically-selected quasars as radio-loud (Schneider {et~al.} 1992).  This corresponds to
$11.8 \pm 5.5$\%\ of the quasars being radio-loud.  No evolution in the
radio-loud fraction is detected.

The FIRST/NVSS detections of optically-selected quasars at $z > 4$ also
provides a lower limit to the radio-loud fraction at early cosmic
epoch.  Of the 107 $z > 4$ optically-selected quasars, 79 have optical
luminosities $-26 > M_B > -28$.  From this restricted sample, 35
overlap with FIRST of which one was detected, implying a statistically
unrobust radio-loud fraction $> 3$\%.  All 79 quasars in the
luminosity range considered overlap with NVSS; four were detected,
implying a radio-loud fraction $> 5$\% at $z > 4$.

\section{Conclusions}

We report on two programs to identify radio-loud quasars at
high-redshift.  First, we correlate the 1.4~GHz FIRST survey with the
digitized 2$^{nd}$ generation Palomar Observatory Sky Survey.
Targeting red stellar optical identifications of radio sources, we
identify six new high-redshift, radio-loud quasars, including three at
$z > 4$.  Next, we consider deep, targeted 4.8~GHz imaging of 32 $z >
4$ quasars selected from the Palomar multicolor quasar survey.  We also
correlate a comprehensive list of 134 $z > 4$ quasars, entailing all
such sources we are aware of as of mid-1999, with two deep 1.4~GHz VLA
sky surveys.  In total, these projects find eight new $z > 4$
radio-loud quasars.

We use this new census to probe the evolution of the radio-loud
fraction with redshift.  We find that, for $-25 \simgt M_B \simgt -28$
and $2 \simlt z \simlt 5$, radio-loud fraction is independent of
optical luminosity.  We also find no evidence for radio-loud fraction
depending on redshift.  If the conventional wisdom that radio-loud AGN
are preferentially identified with early-type galaxies remains robust
at high redshift, this result provides further evidence for an early
formation epoch for elliptical galaxies.  In hierarchical models of
galaxy formation, one expects late-type (less massive) systems to
form first.  Early-type hosts, produced from the merger of multiple
less-massive systems, will have a delayed entrance onto the cosmic
stage with a temporal delay set by the galaxy merger time scale.

\bigskip 
\bigskip 

I thank Hans Hippelein and Klaus Meisenheimer for organizing and hosting
this exceptional meeting at a exceptional site:  the Alpine vista only
marginally competes with the Hexenzimmer for magnificence and splendor.
I wish also to acknowledge the efforts of the POSS-II and DPOSS teams,
which produced the PSS sample used in this work, and, in particular,
N. Weir, J.  Kennefick, R. Gal, S. Odewahn, R. Brunner, V. Desai, and J.
Darling.  The DPOSS project is supported by a generous grant from the
Norris Foundation.  The VLA of the National Radio Astronomy Observatory is
operated by Associated Universities, Inc., under a cooperative agreement
with the National Science Foundation.



\begin{thebibliography}

\bibitem{}{}{} Becker, R.~H., Gregg, M.~D., Hook, I.~M., McMahon, R.~G., White, R.~L., \&  Helfand, D.~J. 1997, \apjl, 479, 93

\bibitem{}{}{} Becker, R.~H., White, R.~L., \& Helfand, D.~J. 1995, \apj, 450, 559

\bibitem{}{}{} Condon, J.~J., Cotton, W.~D., Greisen, E.~W., Yin, Q.~F., Perley, R.~A.,  Perley, G.~B., Taylor, G.~B., \& Broderick, J.~J. 1998, \aj, 115, 1693

\bibitem{}{}{} Djorgovski, S.~G., Gal, R.~R., Odewahn, S.~C., de~Calvalho, R.~R., Brunner, R.,  Longo, G., \& Scaramella, R. 1999, in {\it Wide Field Surveys in Cosmology},  ed. Y.~Mellier \& S.~Colombi (Gif sur Yvette: Editions Fronti\`eres), 89

\bibitem{}{}{} Djorgovski, S.~G., Pahre, M.~A., Bechtold, J., \& Elston, R. 1996, \nat, 382,  234

\bibitem{}{}{} Goldschmidt, P., Kukula, M.~J., Miller, L., \& Dunlop, J.~S. 1999, \apj, 511,  612

\bibitem{}{}{} Gregg, M.~D., Becker, R.~H., White, R.~L., Helfand, D.~J., McMahon, R.~G., \&  Hook, I.~M. 1996, \aj, 112, 407

\bibitem{}{}{} Henry, J. {et~al.} 1994, \aj, 107, 1270

\bibitem{}{}{} Hook, I.~M. \& McMahon, R.~G. 1998, \mnras, 294, L7

\bibitem{}{}{} Hook, I.~M., McMahon, R.~G., Patnaik, A.~R., Browne, I.~W.~A., Wilkinson,  P.~N., Irwin, M.~J., \& Hazard, C. 1995, \mnras, 273, L63

\bibitem{}{}{} Hooper, E.~J., Impey, C.~D., Foltz, C.~B., \& Hewett, P.~C. 1995, \apj, 445, 62

\bibitem{}{}{} Irwin, M., McMahon, R.~G., \& Hazard, C. 1991, in {\it The Space Distribution  of Quasars }, ed. D.~Crampton, Vol.~21 (San Francisco: ASP Conference  Series), 127

\bibitem{}{}{} Kellerman, K.~I., Sramek, R., Schmidt, M., Shaffer, D.~B., \& Green, R. 1989,  \aj, 98, 1195

\bibitem{}{}{} Kennefick, J.~D., Djorgovski, S.~G., \& de~Calvalho, R.~R. 1995, \aj, 110, 2553

\bibitem{}{}{} Kennefick, J.~D., Djorgovski, S.~G., \& Meylan, G. 1996, \aj, 111, 1816

\bibitem{}{}{} {La~Franca}, F., Georgorini, L., Christiani, S., ter, H.~D.~R., \& Owen, F.  1994, \aj, 666, 666

\bibitem{}{}{} McMahon, R.~G., Omont, A., Bergeron, J., Kreyse, E., \& Haslam, C.~G.~T. 1994,  \mnras, 267, 9

\bibitem{}{}{} Miller, J.~S. \& Stone, R.~P.~S. 1994, Lick Obs. Tech. Reports, 66

\bibitem{}{}{} Miller, L., Peacock, J.~A., \& Mead, A.~R.~G. 1990, \mnras, 244, 207

\bibitem{}{}{} Odewahn, S.~C., Windhorst, R.~A., Driver, S.~P., \& Keel, W.~C. 1996, \apj,  472, 13

\bibitem{}{}{} Oke, J.~B. {et~al.} 1995, \pasp, 107, 375

\bibitem{}{}{} Sandage, A. 1972, \apj, 178, 25

\bibitem{}{}{} Schmidt, M. 1963, \nat, 197, 1040

\bibitem{}{}{} Schmidt, M., Schneider, D.~P., \& Gunn, J.~E. 1995a, \aj, 110, 68

\bibitem{}{}{} Schmidt, M., van Gorkom, J.~H., Schneider, D.~P., \& Gunn, J.~E.  1995b, \aj, 109, 473

\bibitem{}{}{} Schneider, D.~P., Schmidt, M., \& Gunn, J.~E. 1989, \aj, 98, 1507

\bibitem{}{}{} ---. 1997, \aj, 114, 36

\bibitem{}{}{} Schneider, D.~P., van Gorkom, J.~H., Schmidt, M., \& Gunn, J.~E. 1992, \aj,  103, 1451

\bibitem{}{}{} Shaver, P.~A., Hook, I.~M., Jackson, C.~A., Wall, J.~V., \& Kellerman, K.~I.  1999, in {\it Highly-Redshifted Radio Lines}, ed. C.~L. Carilli, S.~J.~E.  Radford, K.~M. Menten, \& G.~I. Langston, Vol. 156 (San Francisco: ASP  Conference Series), 163

\bibitem{}{}{} Shaver, P.~A., Wall, J.~V., \& Kellerman, K.~I. 1996a, \mnras,  278, L11

\bibitem{}{}{} Shaver, P.~A., Wall, J.~V., Kellerman, K.~I., Jackson, C.~A., \& Hawkins,  M.~R.~S. 1996b, \nat, 384, 439

\bibitem{}{}{} Smith, J. {et~al.} 1994, \aj, 108, 1147

\bibitem{}{}{} Stern, D., Djorgovski, S.~G., Perley, R., de~Carvalho, R., \& Wall, J.  2000a, \aj, 132, 1526

\bibitem{}{}{} Stern, D., Odewahn, S.~C., Gal, R., Djorgovski, S.~G., de~Carvalho, R., van  Breugel, W., \& Spinrad, H. 2000b, \aj, in preparation

\bibitem{}{}{} Storrie-Lombardi, L.~J., McMahon, R.~G., Irwin, M.~J., \& Hazard, C. 1996,  \apj, 468, 121

\bibitem{}{}{} Visnovsky, K.~L., Impey, C.~D., Foltz, C.~B., tt, P.~C.~H., Weymann, R.~J., \&  Morris, S.~L. 1992, \apj, 391, 560

\bibitem{}{}{} Webster, R.~L., Francis, P.~J., Peterson, B.~A., Drinkwater, M.~J., \& Fasci,  F.~J. 1994, \nat, 375, 469

\bibitem{}{}{} Wilson, A.~S. \& Colbert, E.~J.~M. 1995, \apj, 438, 62

\bibitem{}{}{} Zickgraf, F.-J., Voges, W., Krautter, J., Thiering, I., Appenzeller, I.,  Mujica, R., \& Serrano, A. 1997, \aap, 323, L21

\end{thebibliography}
\end{document}